\documentclass{ws-ijmpd}
\usepackage{amsmath}
\usepackage{amssymb}

\begin{document}

\markboth{Ulhoa et. al} {The Gravitational Energy Problem for
Cosmological Models in Teleparallel Gravity}

\catchline{}{}{}{}{}

\title{The Gravitational Energy Problem for Cosmological Models in Teleparallel Gravity}
\author{S. C. Ulhoa }
\address{International Institute for Physics,\\
Universidade Federal do Rio Grande do Norte, Campus
Universit\'ario, Lagoa Nova. Natal-RN, Brazil. P. O. Box: 1641
59.072-970. \\
sc.ulhoa@gmail.com}

\author{J. F. da Rocha Neto}
\address{Instituto de F\'isica, Universidade de Bras\'ilia
70910-900,\\
Brasilia, DF, Brazil.\\
rocha@fis.unb.br}

\author{J. W. Maluf}
\address{Instituto de F\'isica, Universidade de Bras\'ilia
70904-970,\\
Brasilia, DF, Brazil.\\
P.O. Box: 04385.\\
wadih@unb.br}

\date{\today}

\maketitle

\begin{history}
\received{Day Month Year}
\revised{Day Month Year}
\comby{Managing Editor}
\end{history}

\begin{abstract}

We present a method to calculate the gravitational energy when
asymptotic boundary conditions for the space-time are not given.
This is the situation for most of the cosmological models. The
expression for the gravitational energy is obtained in the context
of the teleparallel equivalent of general relativity. We apply our
method first to the Schwarzschild-de Sitter solution of Einstein's
equation, and then to the Robertson-Walker Universe. We
show that in the first case our method leads to an average energy
density of the vacuum space-time, and in latter case the energy
vanishes in the case of null curvature.

\end{abstract}



\section{Introduction}
\noindent

Recently some investigations have revived the question about the
total energy of the Universe, including the material energy and the
energy of the gravitational field \cite{Rosen} \cite{Sen}
\cite{Vargas} \cite{CopI} \cite{Cop} \cite{Xulu} \cite{lrr-2009-4}.
This question is nontrivial because the usual treatment of the
energy of a cosmological model with pseudotensors or Komar's
integral does not seem appropriate, since the model is described by
a metric that is not asymptotically flat \cite{VF}. The problem
holds even in an empty Universe described by de Sitter's solution,
since in this case there is a (non-flat) cosmological horizon.

The motivation for finding an expression for the energy of the
universe comes from the fact that this subject may help to answer
some questions about the origin and evolution of the universe, with
consequent improvements for the cosmological models \cite{Cop}.

Several investigations and results indicate that the total energy of
the universe is zero. However these results are dubious since most
of them are obtained by means of pseudotensors, which are
coordinate-dependent expressions whose geometrical meaning is not
clear. The results obtained by means of Komar's integrals, on the other
hand, depend on the normalization of a Killing vector and in the
case of cosmological systems, which are not asymptotically flat,
there is no physically preferred choice for such vectors.

In the framework of the Teleparallel Equivalent of General
Relativity (TEGR) the gravitational energy is well defined for
finite volumes of the three dimensional space, and in particular for
asymptotically flat space-times. When asymptotic boundary condition
are not available, for instance in the case of the Robertson-Walker
model and of the de Sitter Universe, a new method has to be used. As
remarked by Faddeev~\cite{Fad}, an expression for the gravitational
energy must vanish only in the total absence of matter and
gravitational field. Therefore in order to avoid the use of boundary
conditions, it is possible, and perhaps mandatory, to use
regularized expressions~\cite{Maluf:2005sr}, which do not require
the establishment of asymptotic conditions. The purpose of this
paper is to point out that the use of a regularized expression for
the gravitational energy is quite suitable to cosmological models.

This paper is organized as follows. In section 2, we introduce the
formalism of teleparallel gravity. We show how to define the
gravitational energy in this context and present the definition of
the regularized expression for the gravitational energy, which can
be applied for arbitrary field configurations. In section 3, we
propose a method to evaluate the energy based on the regularized
expression, since in cosmology the lack of spatial asymptotic
conditions prevents the application of ordinary methods. We evaluate
the energy for two configurations which have relevance in cosmology,
the Schwarzschild-de Sitter's solution and the Robertson-Walker
model for the universe. Finally we present some concluding remarks.

\bigskip
Notation: space-time indices $\mu, \nu, ...$ and SO(3,1) indices $a,
b, ...$ run from 0 to 3. Time and space indices are indicated
according to $\mu=0,i,\;\;a=(0),(i)$. The tetrad field is denoted by
$e^a\,_\mu$, and the flat, Minkowski space-time metric tensor raises
and lowers tetrad indices and is fixed by $\eta_{ab}=e_{a\mu}
e_{b\nu}g^{\mu\nu}= (-1,+1,+1,+1)$. The determinant of the tetrad
field is represented by $e=\det(e^a\,_\mu)$ and we use the constants
$G=c=1$.\par
\bigskip

\section{The regularized gravitational energy-momentum expression in TEGR}
\noindent

The TEGR is an alternative formulation of the theory of general
relativity, and is constructed out of tetrad fields in the
Weitzenb\"{o}ck space-time of distant parallelism. In the
Hamiltonian formulation the constraint equations are interpreted as
energy, momentum and angular momentum equations for the
gravitational field~\cite{Maluf:2006gu}. Two vectors are said to be
parallel if their projections on the tangent space by means of the
action of the tetrad field are equal. Thus, considering two vectors,
$V^a(x)=e^a\,_\mu V^\mu(x)$ and $V^a(x+dx)=e^a\,_\mu V^\mu(x)+
(e^a\,_\lambda\partial_\mu V^\lambda+V^\lambda\partial_\mu
e^a\,_\lambda)dx^{\mu} =V^a+ e^a\,_\mu(\nabla_\lambda
V^\mu)dx^\lambda$, distant from each other by an infinitesimal
displacement, the teleparallelism or distant parallelism is obtained
if

$$\nabla_\nu V^\mu=\partial_{\nu}V^\mu+
(e^{a\mu}\partial_{\nu}e_{a\lambda})V^\lambda=0\,,$$ where
$e^{a\mu}\partial_{\nu}e_{a\lambda}=\Gamma^\mu_{\nu\lambda}$ is
called the Weitzenb\"{o}ck connection. It is simple to check that
$\nabla_\mu e^a\,_\lambda\equiv 0$.

It is well known that in the Riemannian geometry the Christoffel
symbols ${}^0\Gamma^\lambda_{\mu\nu}$ are symmetric in the lower
indices and the corresponding torsion tensor vanishes. However, the
field equations in teleparallel gravity are constructed out of the
torsion tensor $T^\lambda\,_{\mu\nu}$ of the Weitzenb\"ock
connection, where
$T^\lambda\,_{\mu\nu}=e_a\,^\lambda\,T^a\,_{\mu\nu}$, and

\begin{equation}
T^{a}\,_{\mu\nu}(e)=\partial_\mu e^{a}\,_{\nu}-\partial_\nu
e^{a}\,_{\mu}\,. \label{3.0}
\end{equation}
The curvature tensor constructed out of the Weitzenb\"ock connection
$\Gamma^\lambda_{\mu\nu}=e^{a\lambda}\partial_\mu e_{a\nu}$ vanishes
identically.

Let ${}^0\omega_{\mu ab}$ represent the torsion-free Levi-Civita
connection,

\begin{eqnarray}
^0\omega_{\mu ab}&=&-{1\over 2}e^c\,_\mu(
\Omega_{abc}-\Omega_{bac}-\Omega_{cab})\,, \\ \nonumber
\Omega_{abc}&=&e_{a\nu}(e_b\,^\mu\partial_\mu
e_c\,^\nu-e_c\,^\mu\partial_\mu e_b\,^\nu)\,,\label{3.01}
\end{eqnarray}
The Christoffel symbols ${}^0\Gamma^\lambda_{\mu\nu}$ and the
Levi-Civita connection are identically related by $\partial_\mu
e^a\,_\nu +
{}^0\omega_\mu\,^a\,_b\,e^b\,_\nu-{}^0\Gamma^\lambda_{\mu\nu}e^a\,_\lambda=0$,
or

$$^0\Gamma^\lambda_{\mu\nu}=e^{a\lambda}\partial_\mu e_{a\nu}+
e^{a\lambda}\,(^0\omega_{\mu ab})e^b\,_\nu\,.$$ In the teleparallel
equivalent of GR the following {\it identity} is relevant in the
construction of the Lagrangian density,

\begin{equation}
^0\omega_{\mu ab}=-K_{\mu ab}\,, \label{3.02}
\end{equation}
where $K_{\mu ab}=\frac{1}{2}e_{a}\,^{\lambda}e_{b}\,^{\nu}
(T_{\lambda\mu\nu}+T_{\nu\lambda\mu}+T_{\mu\lambda\nu})$ is the
contorsion tensor.

Let us first write the curvature scalar constructed out of
(\ref{3.01}). It is possible to show that such a quantity is given
by

\begin{equation}
eR(^\circ\omega)\equiv -e\left({1\over 4}T^{abc}T_{abc}+{1\over
2}T^{abc}T_{bac}-T^aT_a\right) +2\partial_\mu(eT^\mu)\,.\label{3.1}
\end{equation}
where $e$ is the determinant of the tetrad field $e^a\,_\mu$ and
$T^a=T^b\,_b\,^a$. Both sides of (\ref{3.1}) are invariant under
Lorentz transformations. Dropping the divergence term we construct
the Lagrangian density

\begin{eqnarray}
{\cal L}(e_{a\mu})&=&-k\,e\,\left({1\over 4}T^{abc}T_{abc}+
{1\over 2} T^{abc}T_{bac} -T^aT_a-2\Lambda\right)-{{\cal L}}_M\nonumber \\
&\equiv&-k\,e \Sigma^{abc}T_{abc} -{{\cal L}}_M
+2ke\Lambda\;,\label{3.2}
\end{eqnarray}
where $k=1/(16 \pi)$, ${{\cal L}}_M$ stands for the Lagrangian
density for the matter fields, $\Lambda$ is the cosmological
constant and $\Sigma^{abc}$ is defined by

\begin{equation}
\Sigma^{abc}={1\over 4} (T^{abc}+T^{bac}-T^{cab}) +{1\over 2}(
\eta^{ac}T^b-\eta^{ab}T^c)\;.\label{3.3}
\end{equation}

Performing the variational derivative of the Lagrangian density with
respect to the tetrad field $e_{a\lambda}$ we get, after some
rearrangements, the field equation

\begin{equation}
\partial_\nu(e\Sigma^{a\lambda\nu})={1\over {4k}}
e\, e^a\,_\mu( t^{\lambda \mu} + T^{\lambda \mu})+{1\over
{2}}\Lambda e e^{a\lambda}\;,\label{3.4}
\end{equation}
where

\begin{equation}
t^{\lambda \mu}=k(4\Sigma^{bc\lambda}T_{bc}\,^\mu- g^{\lambda
\mu}\Sigma^{bcd}T_{bcd})\,,\label{3.5}
\end{equation}
and $e^a\,_\mu T^{\lambda \mu}=\frac{1}{e}\frac{\delta {{\cal
L}}_M}{\delta e_{a\lambda}}$ corresponds to the energy-momentum
tensor of matter fields. It is possible to show, by explicit
calculations, that (\ref{3.4}) is equivalent to the Einstein's field
equations with the cosmological term.

Now let us analyze the meaning of $t^{\lambda \mu}$. In view of the
ant-symmetry property $\Sigma^{a\mu\nu}=-\Sigma^{a\nu\mu}$ it
follows that

\begin{equation}
\partial_\lambda
\left[e\, e^a\,_\mu( t^{\lambda \mu} + T'^{\lambda
\mu})\right]=0\,,\label{3.6}
\end{equation}
which is a local balance equation, where $T'^{\lambda
\mu}=T^{\lambda \mu}+2k\Lambda g^{\lambda \mu}$. Therefore we
identify $t^{\lambda\mu}$ as the gravitational energy-momentum
tensor~\cite{Maluf2}.

The integration of $t^{\lambda \mu} + T'^{\lambda \mu}$ over a
hyper-surface $x^0=constant$ defines the energy-momentum vector due
to gravitational and matter fields,

\begin{equation}
P^a=\int_V d^3x\,e\,e^a\,_\mu (t^{0\mu} +T'^{0\mu})\,,\label{3.8}
\end{equation}
where $V$ is a volume of the three-dimensional space. Since the
integrand is a scalar density under coordinate transformations, $V$
may be finite as well as infinite (in which case $S$ is finite or
$S\rightarrow \infty$, respectively). In view of the field
equations, eq. (\ref{3.8}) may be written as

\begin{equation}
P^a=-\int_V d^3x \partial_j \Pi^{aj}=-\oint_S
dS_j\,\Pi^{aj}\,,\label{3.9}
\end{equation}
where $\Pi^{aj}=-4ke\,\Sigma^{a0j}$. It is interesting to note that
the above expression is invariant under coordinate transformations
of the three-dimensional space, under time reparametrizations, and
is projected on the tangent space which means that it is
coordinate independent. We point out that the definition above was
first presented in Ref. ~\cite{GRG1999}.

In this formalism gravitation is considered as a manifestation of
torsion. Therefore in the flat space-time the torsion tensor is
supposed to vanish. However this condition in some cases is not
satisfied~\cite{Maluf:2005sr}, and in this case we have to adopt
regularized expressions. A regularized expression for the
gravitational energy-momentum~\cite{Maluf:2005sr} is defined by

\begin{equation}
P^a_{reg}=-\int_V d^3x\,\partial_k\lbrack\Pi^{ak}(e) -
\Pi^{ak}(E)\rbrack\;, \label{3.10}
\end{equation}
where $\Pi^{aj}(E)$ is constructed out of {\it flat tetrads}
$E^a\,_\mu$ which are obtained from $e^a\,_\mu$ by requiring the
vanishing of the physical parameters. This definition guarantees
that the energy-momentum of the flat space-time always vanishes. In
the context of the energy problem in general relativity, the idea
and concept of a regularized expression was first considered by
Brown and York ~\cite{BY1,BY} in the realm of the quasilocal
definition of the gravitational energy. Conceptually, our approach
regarding the regularization does not differ from that considered by
these authors.

In eq. (\ref{3.10}) $e^a\,_\mu$ and $E^a\,_\mu$ are separately
solutions of Einstein's equations. The flat tetrads $E^a\,_\mu$ may
yield nonvanishing values for the tensor $T^a\,_{\mu\nu}$, and for
this reason the regularization procedure is necessary. Note that
definition (\ref{3.8}) follows from Einstein's equations, and a
similar equation holds for $E^a\,_\mu$. Thus eq. (\ref{3.10}) may be
understood as the subtraction of the two Einstein's equations for
$e^a\,_\mu$ and $E^a\,_\mu$. We remark, however, that in view of eq.
(\ref{3.1}) a local SO(3,1) transformation of the tetrad field
yields only a surface term in the Lagrangian density (\ref{3.2}).
The surface term does not contribute to the field equations, which
are unaffected by a local SO(3,1) transformation of the tetrad
fields.

Tetrad fields that in the flat space-time limit reduce to unit
vectors in the $r$, $\theta$ and $\phi$ directions yield
nonvanishing torsion tensor components ~\cite{Maluf:2005sr}. The
great advantage of using such tetrad fields is that they do not
require the prescription of boundary conditions. This is precisely
the case in cosmology. However, they require the use of a
regularization procedure.

\section{The Procedure to Evaluate the Gravitational Energy}
\noindent

For any solution of Einstein's field equations, such as the
Schwarzschild or Kerr solutions, the choice of the tetrad field
fixes the frame adapted to a certain observer in space-time. Such a
choice may be motivated, for instance, by the asymptotic conditions
of the tetrad at spatial infinity, in which case we take the tetrad
field to represent the flat space-time in this limit. However, in
cosmology this is not the usual situation, since one does not
require boundary conditions for the metric tensor as one does in the
analysis of localized material systems in gravitation. Instead of
boundary conditions, in cosmology one usually requires the concept
of homogeneity of the three-dimensional space, in agreement with the
{\it cosmological principle}, which asserts that we do not occupy
any special position in the universe. The concept of isotropy of
space is an additional requirement of the cosmological models. Thus
in cosmology we consider the tetrad field to represent a local basis
for an arbitrary observer in the chosen coordinate system, and make
use of the regularization procedure. Let us analyze the procedure in
the following cases.

\subsection{The Schwarzschild-de Sitter Space-Time}
\noindent

The most general spherically symmetric vacuum solution of the field
equations with a positive cosmological constant $\Lambda$ is the
Schwarzschild-de Sitter solution,

\begin{equation}
ds^2=-\left(1-\frac{2m}{r}-\frac{r^2}{R^2}\right)dt^2+
\left(1-\frac{2m}{r}-\frac{r^2}{R^2}\right)^{-1}dr^2+r^2
d\theta^2+r^2\sin^2\theta d\phi^2\,,\label{4.1}
\end{equation}
where $R=\sqrt{{3\over \Lambda}}$. This metric specifies the de
Sitter vacuum solution when we set $m=0$, and the Schwarzschild
solution when $\Lambda=0$.

Let us choose a diagonal tetrad field like

\begin{equation}
e^{a}\,_{\mu}(t,r,\theta,\phi)=\left(
                \begin{array}{cccc}
                  A & 0 & 0 & 0 \\
                  0 & A^{-1} & 0 & 0 \\
                  0 & 0 & r & 0 \\
                  0 & 0 & 0 & r\sin\theta \\
                \end{array}
              \right)\,, \label{4.2}
\end{equation}
where $A=\left(1-\frac{2m}{r}-\frac{r^2}{R^2}\right)^{1/2}$. The
determinant $e$ of (\ref{4.2}) reads $e=r^2\sin\theta$. In terms of
one-forms the spatial components are given by $e^{(1)}=A^{-1}dr$,
$e^{(2)}=r\,d\theta$ and $e^{(3)}=r\,\sin\theta\, d\phi$. The
spatial components of the tetrad field represent unit vectors in the
$r$, $\theta$ and $\phi$ directions. Therefore such a tetrad field
is interpreted as the usual basis in spherical coordinates for local
observers.

In order to calculate the energy of the configuration defined by
(\ref{4.1}), we have to evaluate the component $\Sigma^{001}$,
obtained from (\ref{3.3}), which after algebraic manipulations is
given by

\begin{equation}
\Sigma^{001}=\frac{1}{2}g^{00}g^{11}(g^{22}T_{212}+g^{33}T_{313})\,,
\label{4.3}
\end{equation}
where the components of torsion tensor appearing in the above
expression are

\begin{eqnarray}
T_{212}&=&r\,,\nonumber\\
T_{313}&=&r\sin^2\theta\,.\label{4.4}
\end{eqnarray}

Thus integrating over a surface of constant radius $r_0$ in
(\ref{3.9}) it follows that

\begin{eqnarray}
P^{(0)}&=&4k\oint_S dS_1 (e\Sigma^{001}) \nonumber\\
&=&\lim_{r \rightarrow r_0}( -r\sqrt{-g_{00}})\,. \label{4.5}
\end{eqnarray}
However, using the regularized expression given by eq. (\ref{3.10}),
we find that the energy contained within a surface of constant
radius $r_0$ is

\begin{equation}
P^{(0)}_{reg}= r_0\left( 1-\sqrt{
1-\frac{2m}{r_0}-\frac{r_0^2}{R^2}} \right)\,. \label{4.6}
\end{equation}

Let us evaluate expression (\ref{4.6}) for the range of values of
$r_0$ such that

$${{2m}\over {r_0}}\ll 1\;\;,\;\;{{r_0}^2\over {R^2}}\ll 1\;.$$

The above conditions mean that the cosmological constant is very
small. Expanding (\ref{4.6}) and neglecting all powers of both
${{2m}\over {r_0}}$ and ${{r_0}^2\over {R^2}}$ we arrive at

\begin{equation}
P^{(0)}_{reg}=m + {{r_0^3}\over 6}\Lambda\,.\label{4.7}
\end{equation}
which yields, for $\Lambda=0$, the well known value of the energy of
the Schwarzschild space-time.

Now we will instead consider the vacuum de Sitter solution only,
since in this case the main features are not altered by the
introduction of a mass $m$ at $r=0$. Let us assign $E_{dS}$ as the
value of the energy in the absence of the mass $m$, which is
the background (vacuum) energy.

The total gravitational energy $E_{dS}$ contained in the physical
region of the vacuum de Sitter space is obtained from (\ref{4.6}) by
making $m=0$ and $r_0=R$. It follows that

\begin{equation}
E_{dS}=R=\sqrt{\frac{3}{\Lambda}}\,.\label{4.8}
\end{equation}
If we substitute $r = R\sin \chi$ in (\ref{4.1}), then we shall have
a space where the coordinate $\chi$ varies from $0$ to $\pi$. The
surface of a sphere in these coordinates is $S = 4\pi
R^2\sin^2\chi$, which increase reaching its maximum value $4\pi R^2$
at $\chi = \pi/2$, after which it decrease to zero at $\chi =
\pi$~\cite{Dinverno}. The volume is $V=\int_0^\pi d\chi4\pi
R^3\sin^2\chi$ therefore the mean energy density is given by

\begin{equation}
\frac{E_{dS}}{V}=\frac{E_{dS}}{2\pi^2
R^3}=\frac{\Lambda}{6\pi^2}\,.\label{4.9}
\end{equation}
Thus when we consider an Universe filled with matter, the
accelerated rate of expansion of the Universe, which is usually
explained by the hypothesis  of the so called dark energy, may be
due to the mean energy density of the vacuum space-time. Expression
(\ref{4.9}) already appeared in Ref. ~\refcite{Maluf:2003im}, however
we decide to reconsider the same problem since in the present
procedure we do not specify any boundary conditions for the tetrad
field. This method is better suited to describe cosmological models.

\subsection{The Homogeneous and Isotropic Universe}
\noindent

The cosmological principle asserts that the large-scale structure of
the Universe reveals homogeneity and isotropy~\cite{Landau}. The
most general form of a line element that preserves such features may
be written as~\cite{Dinverno},

\begin{equation}
ds^2 = -dt^2 + a^2(t)\biggl[{dr^2 \over 1-k'r^2} + r^2(d\theta^2 +
\sin^2\theta d\phi^2) \biggr]\,,\label{5}
\end{equation}
where $a(t) = S(t)/|K|^{1/2}$ if $K \ne 0$ and $a(t) = S(t)$ if $K =
0$. $S(t)$ is the scale factor and $K$ is the constant curvature of
space. Here $K=|K|k'$ where $k'$ assumes the values $+1, 0, -1$ which
correspond to (i) a three-space of constant positive curvature, (ii)
a flat space or (iii) a space of constant negative curvature,
respectively. By introducing a new radial parameter $\bar{r}$
related to $r$ by

\begin{equation}
r = \frac{\bar{r}}{1 + k'\bar{r}^2/4}\label{5.001}
\end{equation}
we can write the line element above as

\begin{equation}
ds^2 = -dt^2 + \frac{[a(t)]^2}{(1 +
k'\bar{r}^2/4)^2}\left(d\bar{r}^2 + \bar{r}^2(d\theta^2 +
\sin^2\theta d\phi^2)\right)\,.\label{5.002}
\end{equation}
This second form is called the Roberston-Walker (RW) line element.
In the following, we will consider the first line element. The
quantity $a(t)$ is obtained as a solution of the Friedman's
equations,
\begin{eqnarray}
&&3{\dot{a}^2\over a^2} + 3{k'\over a^2} = 8\pi \rho\,,\label{5.1}\\
&&{\dot{a}^2\over a^2} + 2{\ddot{a}\over a^2} + {k'\over a^2} =-8\pi
p\,,\label{5.01}
\end{eqnarray}
where $\rho$ and $p$ denote the density of the cosmological medium
and its pressure.

In what follows we will show that in the context of the TEGR it is
possible to arrive at an expression for the gravitational energy of
the universe for closed, open and flat cases, respectively, using
the following diagonal tetrad field,

\begin{equation}
e^{a}\,_{\mu}(t,r,\theta,\phi)=\left(
                \begin{array}{cccc}
                  1 & 0 & 0 & 0 \\
                  0 & A & 0 & 0 \\
                  0 & 0 & ra(t) & 0 \\
                  0 & 0 & 0 & ra(t)\sin\theta \\
                \end{array}
              \right)\,, \label{5.2}
\end{equation}
where $A= \frac{a(t)}{\sqrt{(1-k'r^2)}}$. The determinant $e$ of
(\ref{5.2}) reads $e=Aa^2(t)r^2\sin\theta$.

In order to obtain the energy of this system, we have to calculate
$\Sigma^{001}$, which reads

\begin{equation}
\Sigma^{001}=\frac{1}{2}g^{00}g^{11}(g^{22}T_{212}+g^{33}T_{313})\,,
\label{5.3}
\end{equation}
where

\begin{eqnarray}
T_{212}&=&ra^2(t)\,,\nonumber\\
T_{313}&=&ra^2(t)\sin^2\theta\,.\label{5.4}
\end{eqnarray}

Taking into account eq. (\ref{3.9}), we integrate now expression
(\ref{5.3}) over a two-dimensional spacelike surface S defined by a
radius $r = r_0 = constant$. We arrive at,

\begin{eqnarray}
P^{(0)}&=&4k\oint_S dS_1 (e\Sigma^{001}) \nonumber\\
&=&-ar_0\sqrt{1 - k'(r_0)^2}\,.\label{5.5}
\end{eqnarray}
We apply now eq. (\ref{3.10}), and after the regularization
procedure we obtain

\begin{equation}
P^{(0)}_{reg} = ar_0\left(1 - \sqrt{1 -
k'(r_0)^2}\right).\label{5.6}
\end{equation}
The expression (\ref{5.6}) was obtained in Ref. ~\cite{nester} by
means of a totally different method. This fact supports our own
procedure and result.

In the case of flat space-time $k' = 0$, the energy is zero. This
result is consistent with the fact that in this case we have an open
(infinit) space with null curvature. This means that in this
configuration, the gravitational energy is minus the energy of the
fields of matter for any value of volume considered. In this
configuration, the gravitational energy and the energy of matter
fields exactly cancel out and the resultant space is flat.

For the case of $k' = 1$ we have a closed space with constant
positive curvature, and the total energy given by eq. (\ref{5.6})
for an arbitrary volume of space is always positive and linear in
the scale factor. This result reveals the fact that, in this case,
the gravitational energy and the energy of the matter fields do not
exactly cancel out, and the difference is likely to be responsible
for the nonvanishing positive curvature.

Finally in the case of space with constant negative curvature $k' =
-1$ (open space) we can see from eq. (\ref{5.6}) that the resultant
energy (gravitational energy plus the energy of matter fields) is
always negative for any value of volume, in particular it is
infinite negative when $r_0\rightarrow \infty$. Here it is important
to note that in this case we have an infinite space with constant
negative curvature. It is likely that in this case, the total
negative energy, is responsible for the constant negative curvature
of the space.

Recently~\cite{NV} negative energy was obtained for cosmological
models describing open universes. In~\refcite{NV} all 9 types of
Bianchi homogeneous cosmological models has been examined by using
an energy expression obtained out of a Hamiltonian boundary term in
the context of tetrad-teleparallel gravity. In contrast to our
results, the authors have found zero energy for some non-flat
cosmological models. However for space with negative curvature the
results in~\refcite{NV} have the same sign of our result for space
with negative curvature. As stated in~\refcite{NV}, for the case $k'
= -1$, each region of constant negative curvature acts like a
concave lens exactly  as if it had a negative matter density
repelling light rays from these.

\section{Conclusion}
\noindent

It is well known that solutions of Einstein's equations that
describe cosmological models do not have well defined boundary
conditions in the same way as the configurations of isolated
sources. The absence of such conditions in  these configurations is
a difficulty for the determination of a definite expression of
gravitational energy for cosmological models. In this work we have
proposed a coordinate independent method to investigate the energy
of gravitational field configurations when asymptotically flat
conditions are not available. The definition given by eq.
(\ref{3.10}) in the absence of matter fields gives the gravitational
energy, and in the presence of matter fields gives the total energy.

We applied our method to the homogeneous and isotropic Universe and
to the Schwarzschild-de Sitter solution. When the cosmological
constant is set to zero in the latter case, it is well known that
the space-time is asymptotically flat. When the cosmological
constant is nonvanishing, the cosmological horizon prevents the
existence of such boundary (asymptotic) conditions. When one tries
to use the tetrad field as a local frame such that the spatial
components are unit vectors along the $r$, $\theta$ and $\phi$
directions, which is realized by means of a diagonal tetrad of the
type given by eq. (\ref{4.2}), some drawbacks may arise. For
example, the emergence of non-vanishing torsion components in
Minkowski space-time. These drawbacks are eliminated by means of the
regularization procedure, which is similar to that given in the work
by Brown and York~\cite{BY} for quasi-local energy expression. The
regularization procedure is powerful because it dispenses the use of
boundary conditions for the gravitational field.

\bibliography{ref}
\bibliographystyle{unsrt}

\end{document}